\documentclass[usenatbib]{mn2e}
\usepackage{natbib}
\usepackage{epsf}
\usepackage{graphicx}

\title[Polarised infrared X-ray binary jets]{Polarised infrared emission from X-ray binary jets\thanks{Based on observations collected at the European Southern Observatory, Chile, under ESO Programme IDs 076.D-0497 and 275.D-5062.}}
\author[D. M. Russell et al.]
{David M. Russell$^{1,2}$ and Rob P. Fender$^1$
\\ $^1$School of Physics \& Astronomy, University of Southampton, Highfield, Southampton, SO17 1BJ, UK
\\ $^2$Astronomical Institute `Anton Pannekoek', University of Amsterdam, Kruislaan 403, 1098 SJ Amsterdam, the Netherlands
\\ davidr@science.uva.nl, rpf@phys.soton.ac.uk
}

\def\simlt{\mathrel{\rlap{\lower 3pt\hbox{$\sim$}}
        \raise 2.0pt\hbox{$<$}}}
\def\simgt{\mathrel{\rlap{\lower 3pt\hbox{$\sim$}}
        \raise 2.0pt\hbox{$>$}}}

\begin{document}
\maketitle

\begin{abstract}
Near-infrared (NIR) and optical polarimetric observations of a selection of X-ray binaries are presented. The targets were observed using the Very Large Telescope and the United Kingdom Infrared Telescope. We detect a significant level (3$\sigma$) of linear polarisation in four sources. The polarisation is found to be intrinsic (at the $> 3\sigma$ level) in two sources; GRO J1655--40 ($\sim 4$--7\% in $H$ and $Ks$-bands during an outburst) and Sco X--1 ($\sim 0.1$--0.9\% in $H$ and $K$), which is stronger at lower frequencies. This is likely to be the signature of optically thin synchrotron emission from the collimated jets in these systems, whose presence indicates a partially-ordered magnetic field is present at the inner regions of the jets. In Sco X--1 the intrinsic polarisation is variable (and sometimes absent) in the $H$ and $K$-bands. In the $J$-band (i.e. at higher frequencies) the polarisation is not significantly variable and is consistent with an interstellar origin. The optical light from GX 339--4 is also polarised, but at a level and position angle consistent with scattering by interstellar dust. The other polarised source is SS 433, which has a low level (0.5--0.8\%) of $J$-band polarisation, likely due to local scattering. The NIR counterparts of GRO J0422+32, XTE J1118+480, 4U 0614+09 and Aql X--1 (which were all in or near quiescence) have a linear polarisation level of $< 16$\% (3$\sigma$ upper limit, some are $< 6$\%). We discuss how such observations may be used to constrain the ordering of the magnetic field close to the base of the jet in such systems.

\end{abstract}


\section{Introduction}

Infrared polarimetric studies of low-mass X-ray binaries (LMXBs; where a compact object -- a black hole or a neutron star -- accretes from a companion star due to Roche lobe overflow) are few and far between. However, such studies can provide information about the physical conditions of the system and inner accretion flow. Most radiation from LMXBs is expected to be unpolarised, for example thermal blackbody radiation from the accretion disc or companion star. The scattering of unpolarised photons could result in a small degree of net polarisation in certain geometries \citep[e.g.][]{dola84}. There is one emission mechanism present in LMXBs that intrinsically produces polarised light -- synchrotron emission. It has been known for decades that optically thin synchrotron radiation can produce a high level (tens of percent) of linear polarisation if the magnetic field structure is ordered \citep[e.g.][]{west59,bjorbl82}.

Linear polarisation (LP) of black hole X-ray binaries (BHXBs) is measured at radio frequencies at a level of $\sim 1$--3\% in a number of sources in an X-ray low/hard state, and up to $\sim 30$\% during transient radio events associated with X-ray state transitions and jet ejections (for a review see \citealt{fend06} and see e.g. \citealt{homabe05} and \citealt{mcclet06} for the definitions of X-ray states). The radio emission of LMXBs is synchrotron in nature and ubiquitously originates in their collimated jets in both black hole and neutron star systems \citep[e.g.][]{fend06,miglfe06}. During transient jet ejections the synchrotron spectrum is optically thin, with a negative spectral index $\alpha$ (where $F_{\nu} \propto \nu^{\alpha}$). For optically thin synchrotron emission, a strong LP signal is expected, of order 70\% $\times$ $f$, where $f$ can be considered to crudely parameterise the degree of ordering of the large scale magnetic field \citep{rybili79,bjorbl82}. The high polarisation levels measured in the radio from these optically thin ejections have indicated that $f$ may be as large as 0.5, and have provided clues towards the fundamental jet and magnetic field properties \citep{fendet99,hannet00,gallet04,brocet07}.

In the low/hard state of BHXBs the radio jet spectrum is optically thick to relatively large distances (light hours) from the launch region, with $\alpha \sim 0$ \citep[e.g.][]{fend01}. This component appears to extend to the infrared regime, where it breaks to an optically thin spectrum, with $\alpha \sim -0.6$ \citep{corbfe02,buxtba04,homaet05a,russet06,hyneet06}. This optically thin component has also recently been identified in a number of low-mass neutron star X-ray binaries (NSXBs) in the infrared \citep*{miglet06,russet07}. The higher frequency photons from the jet are emitted in its inner regions, closest to the compact object (in the absence of shocks downstream). In the optical regime the reprocessed light from the X-ray illuminated accretion disc usually dominates which, like the optically thick jet, should be no more than $\sim 1$\% polarised. In the frequency range in between, which includes the $JHK$-bands, we expect a strong polarised signal from the optically thin synchrotron emission. It is currently uncertain whether the jet component dominates the near-infrared (NIR) at low luminosities \citep[quiescence; e.g.][]{russet06}. In most systems the companion star comes to dominate \citep[e.g.][]{charco06} but in some it does not \citep*[e.g. GX 339--4;][]{shahet01}. If the value of $f$ is high, a polarised signal from the jet component should be detectable. Higher levels of LP (typically $\sim 5$\% but up to $\sim 20$\% in SS 433), which are variable, have been detected in the ultraviolet \citep[UV;][]{woliet96,dolaet97} and result from a combination of Thompson and Rayleigh scattering.

According to theoretical arguments, the signature of the large-scale magnetic field structure at the base of the jet could be inferred via its polarisation signal \citep[e.g.][]{koidet02}. Radio polarimetry has shown that in some cases, the magnetic field is fairly ordered in the optically thin transient jet ejections. When the radio spectrum is optically thick, LP is detected at a level of a few percent. Consequently, if a high level of LP is observed from the low/hard state infrared optically thin spectrum, the magnetic field at the base of the \emph{steady} low/hard state jet must also be ordered. Therefore infrared LP is key to understanding the conditions of the inner regions of the steady jet flow. In addition, the higher frequency infrared photons do not suffer from Faraday rotation, which can confuse radio results.

In 2006, \citeauthor{dubuch06} were the first to report infrared polarimetric observations of LMXBs. They found no evidence for intrinsic LP in H1743--322 in outburst or GRO J1655--40 in quiescence, but did find significant (at the 2.5$\sigma$ level) LP (which is probably intrinsic) in XTE J1550--564 during a weak X-ray outburst. No polarised standard star was observed so the authors were unable to calibrate the polarisation position angle (PA), however if the calibration correction is small then PA $\sim 10^\circ$, which is perpendicular to the known jet axis \citep{corbet02}. For optically thin synchrotron emission, the polarisation PA is a measure of the electric vector, which is perpendicular to the magnetic field vector. Therefore in XTE J1550--564 the magnetic field may be parallel to the jet. Very recently, \cite{shahet07} performed infrared spectropolarimetry of three LMXBs and found two of them (the NSXBs Sco X--1 and Cyg X--2) to be intrinsically polarised, with an increasing LP at lower frequencies. They interpret this as the first detection of the polarised inner regions of the jets.

In the optical regime of LMXBs, just two sources (A0620--00 and GRO J1655--40) possess intrinsic LP to our knowledge \citep{dolata89,glioet98}. The LP varies as a function of orbital phase and is likely caused by the scattering of intrinsically unpolarised thermal emission \citep[e.g.][]{dola84}. No intrinsic LP has been detected from optical observations of NSXBs, except for tentatively in Aql X--1 \citep{charet80}.

We note that polarimetric observations of jets from Active Galactic Nuclei (AGN; which are resolved and are of course orders of magnitude larger and more powerful than X-ray binary jets) have revealed a strong link between the local magnetic field and the dynamics of the jet. LP levels of $>20$\% are observed in the optical and radio, confirming the emission is synchrotron, and the levels and position angles are often correlated with intensity and morphology \citep[for overviews see][]{saiksa88,perlet06}. In addition, NIR flares from the black hole at the centre of our Galaxy, Sgr A$^\ast$, are also highly polarised ($\sim 10$--20\%) and may originate in its jets \citep{eckaet06,meyeet06}.

Here, we have obtained linear polarimetric observations of eight X-ray binaries using the 8 m Very Large Telescope (VLT) and the 3.8 m United Kingdom Infrared Telescope (UKIRT) in order to constrain the origin of the emission and the level of ordering of the jet magnetic field. Seven of the eight targets are observed in NIR filters and the other, in optical filters. In Section 2 we discuss the target selection, observations, data reduction and polarimetry calibration. The results are discussed in detail in Section 3 and in Section 5 we summarise our results.

\section{Observations}

In an effort to identify the polarised signature of jets in the infrared, we first selected the best suited sources and telescopes. The evidence for optically thin emission from the jet in the NIR (and hence our best chance of detecting high levels of LP) comes mainly from BHXBs in outburst in the low/hard state \citep[e.g.][]{homaet05a,hyneet06} and low-magnetic field NSXBs that are active \citep{miglet06,russet07}. However it has been proposed that the jet may also contribute or dominate the NIR flux in some quiescent BHXBs \citep{russet06} but probably not in quiescent NSXBs \citep{russet07}. Consequently, we obtained observations of both outbursting and quiescent sources.

\subsection{Target selection}

\begin{table*}
\caption{Log of Observations.}
\begin{tabular}{lllllllllll}
\hline
Source&Instrument&Date&MJD&Source activity&\multicolumn{3}{c}{Exposure times (sec)}&Airmass\\
\hline
      &	         &    &   &	          &$V$&$R$&$I$                             &       \\
GX 339--4    &FORS1&22 Aug 2005&53604.1&low luminosity state&264&264&264&1.19--1.26\\
GX 339--4    &FORS1&24 Aug 2005&53606.0&low luminosity state&264&264&264&1.14--1.18\\
GX 339--4    &FORS1&31 Aug 2005&53613.0&low luminosity state&264&264&264&1.13--1.17\\
GX 339--4    &FORS1&12 Sep 2005&53625.0&low luminosity state&264&264&264&1.16--1.23\\
GX 339--4    &FORS1&19 Sep 2005&53632.0&low luminosity state&264&264&264&1.20--1.28\\
Hiltner 652  &FORS1&19 Sep 2005&53632.0&polarised standard  &2&2&2&1.03--1.04\\
\hline
      &	         &    &   &	          &$J$&$H$&$Ks$                            &       \\
GRO J1655--40&ISAAC&14 Oct 2005&53657.0&hard state&&56&&1.52\\
GRO J1655--40&ISAAC&28 Oct 2005&53671.0&hard state&32&56&80&2.05--2.13\\
WD 2359--434 &ISAAC&28 Oct 2005&53671.0&unpolarised standard&14&14&14&1.16--1.17\\
\hline
      &	         &    &   &	          &$J$&$H$&$K$                             &       \\
XTE J1118+480&UIST &15 Feb 2006&53781.5&quiescence&2880&&1440&1.13--1.25\\
Sco X--1     &UIST &15 Feb 2006&53781.6&persistent&3$\times$32&3$\times$28&3$\times$28&1.27--1.48\\
Sco X--1     &UIST &16 Feb 2006&53782.6&persistent&96&84&84&1.37--1.52\\
WD 1344+106  &UIST &17 Feb 2006&53783.5&unpolarised standard&720&&&1.49--1.58\\
XTE J1118+480&UIST &17 Feb 2006&53783.5&quiescence&2880&&&1.13--1.15\\
Sco X--1     &UIST &07 Mar 2006&53801.6&persistent&3$\times$32&3$\times$28&3$\times$28&1.24--1.27\\
SS 433       &UIST &07 Mar 2006&53801.6&persistent&24&&&1.49--1.53\\
HD 38563C    &UIST &08 Mar 2006&53802.3&polarised standard&47&&&1.42--1.48\\
Sco X--1     &UIST &08 Mar 2006&53802.6&persistent&96&84&84&1.37--1.52\\
WD 1615-154  &UIST &08 Mar 2006&53802.6&unpolarised standard&720&&&1.23--1.24\\
SS 433       &UIST &19 Aug 2006&53966.5&persistent&24&&&1.79--1.84\\
GRO J0422+32 &UIST &19 Aug 2006&53966.6&quiescence&2160&&2160&1.17--1.30\\
Aql X--1     &UIST &20 Aug 2006&53967.3&fading from mini outburst&720&&720&1.06--1.10\\
Aql X--1     &UIST &08 Oct 2006&54016.4&quiescence&&&1680&1.51--1.88\\
4U 0614+09   &UIST &08 Oct 2006&54016.7&low, persistent state&&&480&1.02\\
Aql X--1     &UIST &11 Oct 2006&54019.2&quiescence&2160&2160&&1.07--1.21\\
\hline
\end{tabular}
\small
\\ MJD =  Modified Julian Day; Exposure times are total on-source integration time.
\normalsize
\end{table*}

GX 339--4 undergoes quasi-regular outbursts of varying length and peak luminosity. The infrared jet of GX 339--4 is one of the most studied, and the NIR spectrum in the low/hard state is consistent with optically thin synchrotron emission \citep{corbfe02,homaet05a}. This spectral component is observed to join to a thermal spectrum in the optical, with the two components producing half of the flux each around the $I$-band. We were allocated VLT imaging polarimetry with FORS1 (optical) for these observations (ESO Programme ID 076.D-0497). The 2005 outburst of GRO J1655--40 was studied at X-ray, optical and radio wavelengths \citep[e.g.][]{shapet07}. On the outburst decline the source entered the low/hard state \citep{homaet05b} so we obtained NIR imaging polarimetry with a Director's Discretionary Time (DDT) proposal with ISAAC (ESO Programme ID 275.D-5062).

In addition to outbursting sources, we selected a number of quiescent and persistent sources for NIR polarimetry. Target selection of quiescent systems was based on known NIR magnitudes and the companion star contribution, where measured. For example, the companion in the A0620-00 system is known to contribute $\sim 75$\% of the $H$-band quiescent flux \citep*{shahet99}. Even if a polarised jet provides the remaining 25\% (as opposed to the disc), the level of LP would be dampened by a factor of 4 due to the unpolarised light from the star. We found that most NIR-bright BHXBs are dominated by their companion stars \citep[e.g. V404 Cyg;][]{casaet93} but there are some good candidates (e.g. GRS 1915+105; XTE J1118+480). Three NSXBs are also good candidates to identify a NIR jet component. Optically thin synchrotron emission appears to join to the thermal disc spectrum in the NIR in the NSXB 4U 0614+09 \citep{miglet06}. Sco X--1 is the brightest ($J = 11.9$) of the Z-sources, a class of neutron star whose jet spectrum may significantly contribute to the NIR. Aql X--1 is very active (it has about one outburst per year) and has a NIR spectrum in outburst consistent with a jet origin \citep{russet07}.

In Table 1 we list all observations used in this work. For all three instruments (on two telescopes) imaging polarimetry is obtained using a Wollaston prism inserted in the optical path. The light is split into simultaneous, perpendicularly polarised ordinary and extra-ordinary beams (o- and e-beams). The waveplate was rotated to four angles: 0$^\circ$, 22.5$^\circ$, 45$^\circ$ and 67.5$^\circ$; this achieves a higher accuracy of LP and PA than from just two angles 0$^\circ$ and 45$^\circ$. The flux of the source was then measured in the o- and e-beams for each prism angle ($F_0^o$, $F_0^e$, $F_{22}^o$, $F_{22}^e$, etc.) using aperture photometry in \small IRAF \normalsize and LP and PA were calculated using the `ratio' method thus:
\begin{eqnarray}
  R_Q^2 = \frac{(F_0^e/F_0^o)}{(F_{45}^e/F_{45}^o)};~~R_U^2 = \frac{(F_{22}^e/F_{22}^o)}{(F_{67}^e/F_{67}^o)}\\
  q = \frac{R_Q-1}{R_Q+1};~~u = \frac{R_U-1}{R_U+1}\\
  LP = (q^2 + u^2)^{0.5}\\
  PA / deg = 0.5 tan^{-1}(u/q)
\end{eqnarray}

Flux calibration was possible for most of our targets. The total flux of each source was estimated from the sum of the o- and e-beam fluxes at each angle. Magnitudes were converted to de-reddened flux densities by accounting for interstellar extinction. We adopted the method of \cite*{cardet89} and used the known values of $A_{\rm V}$ for each source (Table 2).

\begin{table}
\caption{Values of interstellar extinction towards each source.}
\begin{tabular}{lll}
\hline
Source&$A_{\rm V}$&Reference\\
\hline
GRO J0422+32 &0.74$\pm$0.09&\cite{geliha03}\\
4U 0614+09   &1.41$\pm$0.17&\cite{neleet04}\\
XTE J1118+480&0.053$^{+0.027}_{-0.016}$&\cite{chatet03}\\
Sco X--1     &0.70$\pm$0.23&\cite{vrtiet91}\\
GRO J1655--40&3.7$\pm$0.3&\cite{hyneet98}\\
GX 339--4    &3.9$\pm$0.5&\cite{jonket04}\\
Aql X--1     &1.55$\pm$0.31&\cite{chevet99}\\
SS 433       &6.5$\pm$0.5&\cite{fukuet97}\\
\hline
\end{tabular}
\small
\normalsize
\end{table}

\subsection{ESO VLT observations and reduction}

The VLT FORS1 Service Mode programme 076.D-0497 was carried out between 22 August and 19 September 2005. The target GX 339--4 was observed on five nights in Bessel $V$, $R$ and $I$ filters, and the polarised standard star Hiltner 652, on one night (see Table 1 for observations and Table 3 for the results). The seeing ranged from 0.48 to 2.09 arcsec, with a mean of $\sim 1.2$ arcsec. The conditions were clear on all nights and photometric on three of the five. Using \small IRAF\normalsize, the science images were de-biased and flat-fielded and aperture photometry was performed on GX 339--4 and three stars in the field of view (FOV). The point spread function (PSF) of GX 339--4 was found to be blended with two close stars \citep[B and C in][]{shahet01}. We therefore could not measure small differences in the flux of GX 339--4 alone; we used an aperture which encompassed the PSFs of the contaminating stars. GX 339--4 contributes $\sim 60$\% of the total flux of this aperture during our observations. If our target is polarised and the contaminating stars are not then the flux from them will dampen our LP measurements.

From the background-subtracted photometry we obtained the Stokes parameters $q$ and $u$ (equations 1 and 2), and the variability of the total (o + e) flux across the four prism angles. Instrumental polarisation is documented to be small for FORS1: $0.09\pm 0.01$\% in $V$-band \citep{fosset06}. To calibrate the PA, we make the standard small FORS1 correction of $PA_{\rm REAL} = PA_{\rm MEASURED} + c$, where $c=-1.80^\circ$, $+1.19^\circ$ and $+2.89^\circ$ for the $V$, $R$ and $I$ filters, respectively. For Hiltner 652 we obtained standard FORS1 polarimetric observations in which two prism angles were used as opposed to four (standard recipes require just two, but most of our targets are low S/N and require four angles to achieve higher accuracy of LP). $q$ and $u$ were calculated from these two angles alone for Hiltner 652 (and no variability is quoted). We measure LP $=6.2$\% for this polarised standard in $V$, consistent with that quoted by the FORS consortium based on Commissioning data taken with FORS1. Unfortunately Hiltner 652 was slightly saturated in $R$ and $I$, and our values of LP are lower than expected. Since the known instrumental polarisation is low for FORS1, the V-band measurement is sufficient to insure the instrument is calibrated correctly and our method is correct. The measured PA in all three filters is at most $\sim 3^\circ$ different from the documented values.

No flux standard stars were observed but we estimated the magnitude of GX 339--4 using four stars in the FOV that are listed in the NOMAD and USNO-B Catalogs \citep{zachet04,moneet03}. The magnitudes of these field stars range from $V=14.13$, $I=13.43$ to $V=17.74$, $I=16.39$. Our flux measurements do not require the high level of sensitivity needed for polarimetry, so we estimate the magnitude of GX 339--4 alone (not including the contaminating stars B and C) using a circular aperture of fixed radius centred on GX 339--4. The aperture excludes most of the light from B and C. The 1$\sigma$ errors in the estimated magnitudes of GX 339--4 are inferred from the range of magnitudes implied by each of the four field stars.

\begin{figure}
\centering
\includegraphics[width=6cm,angle=270]{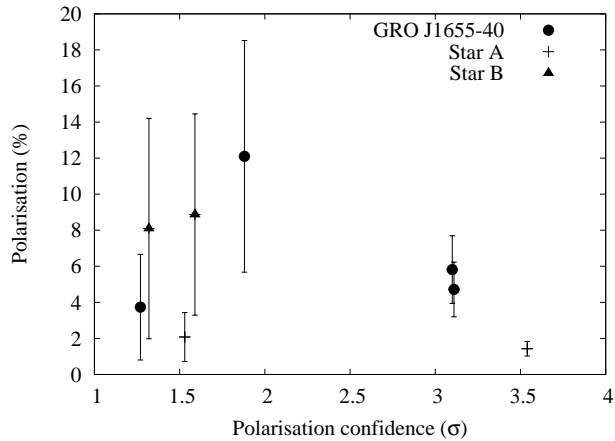}
\caption{Confidence in measured polarisation for GRO J1655--40 and two stars in the same field of view. The only 3$\sigma$ detections of LP are from GRO J1655--40 at a level of $\sim 6$\% (in all observations when the S/N is high) and one of the field stars at $1.4\pm 0.4$\%.
}
\end{figure}

GRO J1655--40 was observed on 14 and 28 October 2005 under VLT ISAAC DDT programme 275.D-5062 (Tables 1 and 3). Short wavelength (SW) imaging polarimetry was carried out with $J$ (1.25 $\mu$m), $H$ (1.65 $\mu$m) and $Ks$ (2.16 $\mu$m) filters. The seeing was 0.71--1.58 arcsec and conditions were photometric on both nights. The unpolarised standard star WD 2359--434 was also observed on 28 October. Data reduction was performed in \small IRAF\normalsize; dark frames of equal exposure time to the science frames were subtracted to remove sky background. The science images were then flat-fielded and bad pixels were accounted for.

We calculated the Stokes parameters (and the variability of the total flux) for GRO J1655--40 and two nearby field stars of similar brightness. In Fig. 1 we plot the measured LP as a function of the confidence of the measurement. The field stars are not polarised (the confidence is $< 2\sigma$) except one observation of $1.4 \pm 0.4$\% LP, which is low and probably interstellar in origin (this is reasonable for the extinction towards GRO J1655--40 of $A_{\rm V}\sim 3.7$; the star may be closer to us than the X-ray binary however). Where the signal-to-noise ratio (S/N) is high enough (Table 3), LP is detected in GRO J1655--40 at a level of 5--6\%, with 3$\sigma$ confidence. The absence of this polarisation in the field stars implies it is intrinsic to GRO J1655--40 and not due to instrumental polarisation. It is also too high to be of interstellar origin because the extinction is too low. Serkowski's law; $A_{\rm V}\geq P(\%)/3$ \citep*{serket75} indicates that the maximum LP caused by interstellar dust in the optical is 12.0\% for GRO J1655--40, adopting the extinction upper limit of $A_{\rm V} < 4.0$ (Table 2). In the $K$-band, $A_{\rm K }\sim 0.1 A_{\rm V}$, so LP$_{\rm K}\leq 1.2$\% -- much lower than is observed. In addition \cite{dubuch06} measured LP$_{\rm K}\leq 1.3$\% during quiescence, so the LP we measure must be transient.

An unpolarised standard, WD 2359--434 was also observed. We detect no LP in WD 2359--434 except a low level ($\sim 2$\%) in the $J$-band, which is likely due to instrumental polarisation. The $J$-band values of $q$ and $u$ for WD 2359--434 were then subtracted from those of GRO J1655--40 to correct for this. No additional systematic correction to the values of PA are required for ISAAC. WD 2359--434 ($J=12.597$, $H=12.427$, $K=12.445$) was also used to calibrate the flux of GRO J1655--40. Atmospheric extinction was accounted for because the standard star and X-ray binary were observed at different airmasses.

\begin{table*}
\caption{List of VLT results.}
\begin{tabular}{lllllllllll}
\hline
Source&MJD&Filter&S/N &app.&$F_{\nu}$&LP (\%)&PA ($^\circ$)&Polarisation		&Variabi-\\
      &   &	 &    &mag.&(mJy)    &	    &                   &confidence ($\sigma$)  &lity (\%)\\
(I)&(II)  &(III) &(IV)&(V) &(VI)     &(VII) &(VIII)             &(IX)                   &(X)\\
\hline
&\multicolumn{9}{c}{\emph{VLT + FORS1 observations$^1$:}}&\\
GX 339--4  &53604.1&$V$&846  &18.80(06) &4.19(24)&2.15$\pm$0.17 &27.4$\pm$4.5   &12.8&3.2$\pm$0.3\\
GX 339--4  &53604.1&$R$&592  &18.70(42)&2.05(79)&2.45$\pm$0.24 &36.4$\pm$5.6   &10.3&1.5$\pm$0.4\\
GX 339--4  &53604.1&$I$&1630 &17.65(21)&2.07(40)&1.78$\pm$0.09 &34.1$\pm$2.8   &20.5&1.9$\pm$0.1\\
GX 339--4  &53606.0&$V$&127  &18.61(17)&5.00(77)&$<5.03$      & 	&    &5.6$\pm$1.8\\
GX 339--4  &53606.0&$R$&80   &18.52(40)&2.38(88)&$<7.42$      & 	&    &26$\pm$3\\
GX 339--4  &53606.1&$I$&95   &17.46(11)&2.47(26)&$<6.59$      & 	&    &29$\pm$2\\
GX 339--4  &53613.0&$V$&237  &18.68(17)&4.66(73)&2.98$\pm$0.60 &23.7$\pm$11.5  &5.0 &6.8$\pm$0.9\\
GX 339--4  &53613.0&$R$&104  &18.67(40)&2.07(76)&$<6.62$      & 	&    &$<11$\\
GX 339--4  &53613.0&$I$&73   &17.41(12)&2.58(27)&$<8.79$      & 	&    &$<14$\\
GX 339--4  &53625.0&$V$&454  &18.56(11)&5.21(54)&2.02$\pm$0.31 &34.3$\pm$8.9   &6.5 &1.6$\pm$0.5\\
GX 339--4  &53625.0&$R$&750  &18.49(40)&2.46(89)&2.86$\pm$0.19 &34.3$\pm$3.8   &15.2&2.3$\pm$0.3\\
GX 339--4  &53625.0&$I$&432  &17.54(15)&2.31(31)&2.33$\pm$0.33 &38.9$\pm$8.0   &7.1 &4.1$\pm$0.5\\
GX 339--4  &53632.0&$V$&659  &18.93(17)&3.71(58)&2.10$\pm$0.21 &24.9$\pm$5.9   &9.8 &2.2$\pm$0.3\\
GX 339--4  &53632.0&$R$&429  &18.87(40)&1.72(64)&2.90$\pm$0.33 &31.3$\pm$6.5   &8.8 &2.7$\pm$0.5\\
\vspace{1mm}
GX 339--4  &53632.0&$I$&355  &17.68(12)&2.02(22)&2.57$\pm$0.40 &27.1$\pm$8.9   &6.5 &2.5$\pm$0.6\\
Hiltner 652    &53632.0&$V$&21000&&&6.19$\pm$0.01&176.2$\pm$0.1&924 &-  \\
Hiltner 652$^2$&53632.0&$R$&32000&&&4.79$\pm$0.01&176.9$\pm$0.1&1088&-\\
Hiltner 652$^2$&53632.0&$I$&26000&&&3.54$\pm$0.01&178.6$\pm$0.1&644 &-\\
\hline
&\multicolumn{9}{c}{\emph{VLT + ISAAC observations:}}&\\
GRO J1655--40&53657.0&$H$&93&12.95(12)&11.8(1.3)&4.72$\pm$1.52&44$\pm$18&3.1&$<11$\\
GRO J1655--40&53671.0&$J$&48&13.76(13)&12.4(1.5)&$< 11.7$&&&$<22$\\
GRO J1655--40&53671.0&$H$&22&13.01(16)&11.2(1.6)&$< 31.4$&&&$<42$\\
\vspace{1mm}
GRO J1655--40&53671.0&$Ks$&75 &12.62(12)&8.04(91)&5.82$\pm$1.88&157$\pm$18&3.1&6.9$\pm$3.0\\
WD 2359--434 &53671.0&$J$ &210&&&2.04$\pm$0.69&153$\pm$19&3.0&$<4.4$\\
WD 2359--434 &53671.0&$H$ &72&&&$< 5.6$&&&$<11$\\
WD 2359--434 &53671.0&$Ks$&56&&&$< 5.4$&&&$<16$\\
\hline
\end{tabular}
\small
$^1$Polarisation measurements of GX 339--4 include the contaminating stars B and C \citep{shahet01} in the aperture;
$^2$Standard star is slightly saturated, which has reduced the apparent LP level detected.
Columns give:
(I) Source name;
(II) Modified Julian Day;
(III) filter (waveband);
(IV) S/N detection of the source;
(V) apparent magnitude;
(VI) de-reddened flux density;
(VII) measured level of polarisation and $1\sigma$ error, or $3\sigma$ upper limit if the detection of polarisation is $< 2\sigma$;
(VIII) polarisation angle and $1\sigma$ error;
(IX) confidence of polarisation detected;
(X) the amplitude of the variability (the standard deviation of the total source intensity as a percentage; $\Delta F_{\nu}$/$F_{\nu}$) and $1\sigma$ error, or $3\sigma$ upper limit if the variability detection is $< 2\sigma$.
\normalsize
\end{table*}

\begin{table*}
\caption{List of UKIRT results$^1$.}
\begin{tabular}{lllllllllll}
\hline
Source&MJD&Filter&S/N &app.&$F_{\nu}$&LP (\%)&PA ($^\circ$)$^2$&Polarisation		&Variabi-\\
      &   &	 &    &mag.&(mJy)    &	    &                   &confidence ($\sigma$)  &lity (\%)\\
(I)&(II)  &(III) &(IV)&(V) &(VI)     &(VII) &(VIII)             &(IX)                   &(X)\\
\hline
GRO J0422+32 &53966.6&$K$&3.5&17.42(0.42)&0.072(28)&            &&   &    \\
\vspace{1mm}
GRO J0422+32 &53966.6&$J$&44&18.36(0.16)&0.083(12)&$< 11.6$     &&   &$<24$\\
\vspace{1mm}
4U 0614+09   &54016.7&$K$&37&16.64(0.19)&0.158(28)&$< 16.0$     &&   &$<20$\\
XTE J1118+480&53781.5&$J$&91&&&$< 6.36$     &&   &$<9.6$\\
XTE J1118+480&53781.6&$K$&66&&&$< 7.42$     &&   &$<15$\\
\vspace{1mm}
XTE J1118+480&53783.5&$J$&28&&&$< 20.7$     &&   &$<28$\\
Sco X--1     &53781.6&$J$&3000--3900&&&0.34--0.53&22--87&9.3--11.5&0.87--1.3\\
Sco X--1     &53781.7&$H$&3000--4500&&&0.30--0.57&27--88&6.4--18.3&1.0--1.9\\
Sco X--1     &53781.6&$K$&1600--2900&&&0.13--0.52&71--86&$<$2--10.7&1.5--4.5\\
Sco X--1     &53782.6&$J$&2000      &&&0.36$\pm$0.07&37$\pm$12&5.0	&0.53$\pm$0.11\\
Sco X--1     &53782.6&$H$&2000      &&&0.18$\pm$0.07&49$\pm$23&2.5	&$<0.54$\\
Sco X--1     &53782.6&$K$&1600      &&&0.23$\pm$0.09&41$\pm$23&2.5	&$<0.50$\\
Sco X--1     &53801.6&$J$&1100--2000&&&0.30--0.65&36--95&2.4--6.9&0.44--1.4\\
Sco X--1     &53801.6&$H$&1200--2000&&&0.07--0.54&36--49&$<$2--7.7&0.45--0.57\\
Sco X--1     &53801.6&$K$&1000--2000&&&0.35--0.91&96--111&2.6--5.0&0.33--0.41\\
Sco X--1     &53802.6&$J$&1600      &&&0.40$\pm$0.09&46$\pm$13&4.4     &0.44$\pm$0.14\\
Sco X--1     &53802.6&$H$&700	    &&&$< 1.01$  &&	   &$<1.5$\\
\vspace{1mm}
Sco X--1     &53802.6&$K$&600	 &&&$< 1.08$	 &&	   &$<1.5$\\
Aql X--1     &53967.3&$J$&88	 &16.30(15)&0.633(87)&$< 5.44$	 &&	   &$<8.4$\\
Aql X--1     &53967.4&$K$&41	 &15.47(16)&0.455(69)&$< 11.6$	 &&	   &$<18$\\
Aql X--1     &54016.4&$K$&28	 &15.56(22)&0.418(85)&$< 16.2$	 &&	   &$<26$\\
Aql X--1     &54019.2&$J$&19	 &16.65(29)&0.459(123)&$< 24.5$	 &&	   &$<37$\\
\vspace{1mm}
Aql X--1     &54019.2&$H$&27	 &16.06(25)&0.451(104)&$< 16.9$	 &&	   &$<26$\\
SS 433       &53801.6&$J$&4600   &9.06(08)&1950(140)&0.75$\pm$0.03&71$\pm$2&24.4        &0.27$\pm$0.05\\
\vspace{1mm}
SS 433       &53966.5&$J$&4000   &8.89(19)&2290(400)&0.50$\pm$0.04&76$\pm$4&14.1        &0.47$\pm$0.06\\
\vspace{1mm}
WD 1344+106  &53783.5&$J$&500    &&&$< 1.08$     &&        &$<1.8$\\
\vspace{1mm}
HD 38563C    &53802.3&$J$&2100   &&&5.56$\pm$0.07&        &84.2        &$<0.47$\\
WD 1615--154 &53802.6&$J$&370    &&&$< 1.79$     &&        &$<2.4$\\
\hline
\end{tabular}
\small
$^1$The very close contaminating star \citep{chevet99} is included in the aperture of Aql X--1.
$^2$The position angle may have systematic errors because it is calibrated using one standard, HD 38563C.
Columns give:
(I) Source name;
(II) Modified Julian Day;
(III) filter (waveband);
(IV) S/N detection of the source;
(V) apparent magnitude;
(VI) de-reddened flux density;
(VII) measured level of polarisation and $1\sigma$ error, or $3\sigma$ upper limit if the detection of polarisation is $< 2\sigma$;
(VIII) polarisation angle and $1\sigma$ error;
(IX) confidence of polarisation detected;
(X) the amplitude of the variability (the standard deviation of the total source intensity as a percentage; $\Delta F_{\nu}$/$F_{\nu}$) and $1\sigma$ error, or $3\sigma$ upper limit if the variability detection is $< 2\sigma$.
\normalsize
\end{table*}

\subsection{UKIRT observations and reduction}

Our UKIRT targets were observed (see Table 1 for observations and Table 4 for results) between 15 February and 11 October 2007 by the UKIRT Service Observing Programme (UKIRTSERV). $JHK$ polarimetric observations were made using UIST + IRPOL2. The conditions were mostly clear during the observations (and the seeing was 0.4--1.0 arcsec) but thin cirrus was present on some dates. The POL\_ANGLE\_JITTER standard recipe was adopted except we did not take flat field images. After dark current and bias subtraction, the flux from the target was measured in a fixed aperture (the same pixels in all images for each source). Since we are dividing the fluxes in these fixed apertures (e.g. $F_0^o/F_{45}^o$; equation 1) the flat fields would cancel out, so we did not require skyflats.

The instrumental polarisation is known to be low ($<1$\%), which is confirmed by the non-detection of LP from the two unpolarised standards WD 1344+106 and WD 1615-154 (Table 4). The PA was calibrated using the polarised standard HD 38563C; we caution that there may be systematic errors in PA due to just the one calibrator being available. No flux standards were observed, however we use relative photometry of two stars in the FOV (sum of o+e beams) that are listed in the 2MASS catalog, for each target. For XTE J1118+480 and Sco X--1, no field stars exist in the images that have 2MASS magnitudes, so no flux calibration could be achieved. We could not use the polarisation standards for flux calibration as conditions were not photometric in all observations.

\section{Results \& Discussion}

We detect a significant level (3$\sigma$) of LP in four sources: GX 339--4, GRO J1655--40, Sco X--1 and SS 433. We find that two of these four (GRO J1655--40 and Sco X--1) have intrinsic LP in the NIR which is most likely caused by optically thin synchrotron emission from the inner regions of the jets. SS 433 is polarised in the NIR due to local scattering and GX 339--4 is polarised in the optical most likely due to interstellar dust along its line of sight. The other four sources are not polarised, with 3$\sigma$ upper limits of $\sim$5--15\% (low S/N prevents us from probing lower levels of LP in most quiescent systems). Here, we discuss the results for each individual source and then rule out alternative origins to the variable LP in Sco X--1 and GRO J1655--40.

\subsection{Individual sources}

\textbf{GX 339--4}:
\newline
We observed this source on five occasions spanning 28 days during a low luminosity state soon after its 2004--5 outburst. LP is detected at a level of 2--3\% in GX 339--4 (which includes the two close contaminating stars, see Section 2.2) in all three optical filters in all data with a high S/N (Table 3). From all the $VRI$ polarisation measurements we obtain a mean LP = 2.4\%. In Fig. 2 we plot LP as a function of PA for GX 339--4 (including the contaminating stars; upper panel) and three other stars in the FOV (lower panel). Both LP and PA for GX 339--4 are similar to those measured from the field stars, with the values of PA differing for a few field star observations. This is a strong indication that the polarisation in all sources (including the contaminating stars B and C) is caused by the same process, namely interstellar dust. Moreover, we do not detect significant variability in LP. The extinction towards GX 339--4 has been measured at $A_{\rm V}=3.9\pm 0.5$ \citep{jonket04} which is high enough to explain the observed polarisation. Indeed, we can constrain the extinction from our measured level of LP using Serkowski's law, $A_{\rm V}\geq P(\%)/3$ \citep{serket75}. We find that $A_{\rm V}\geq 0.81$, which does not refine the currently estimated value of $3.9\pm 0.5$. The axis of the resolved radio jet of GX 339--4 \citep{gallet04} is at 116$^\circ$, approximately perpendicular to the PA of LP in GX 339--4, however there is no indication that the two are linked, for the above reasons.

\begin{figure}
\centering
\includegraphics[width=6cm,angle=270]{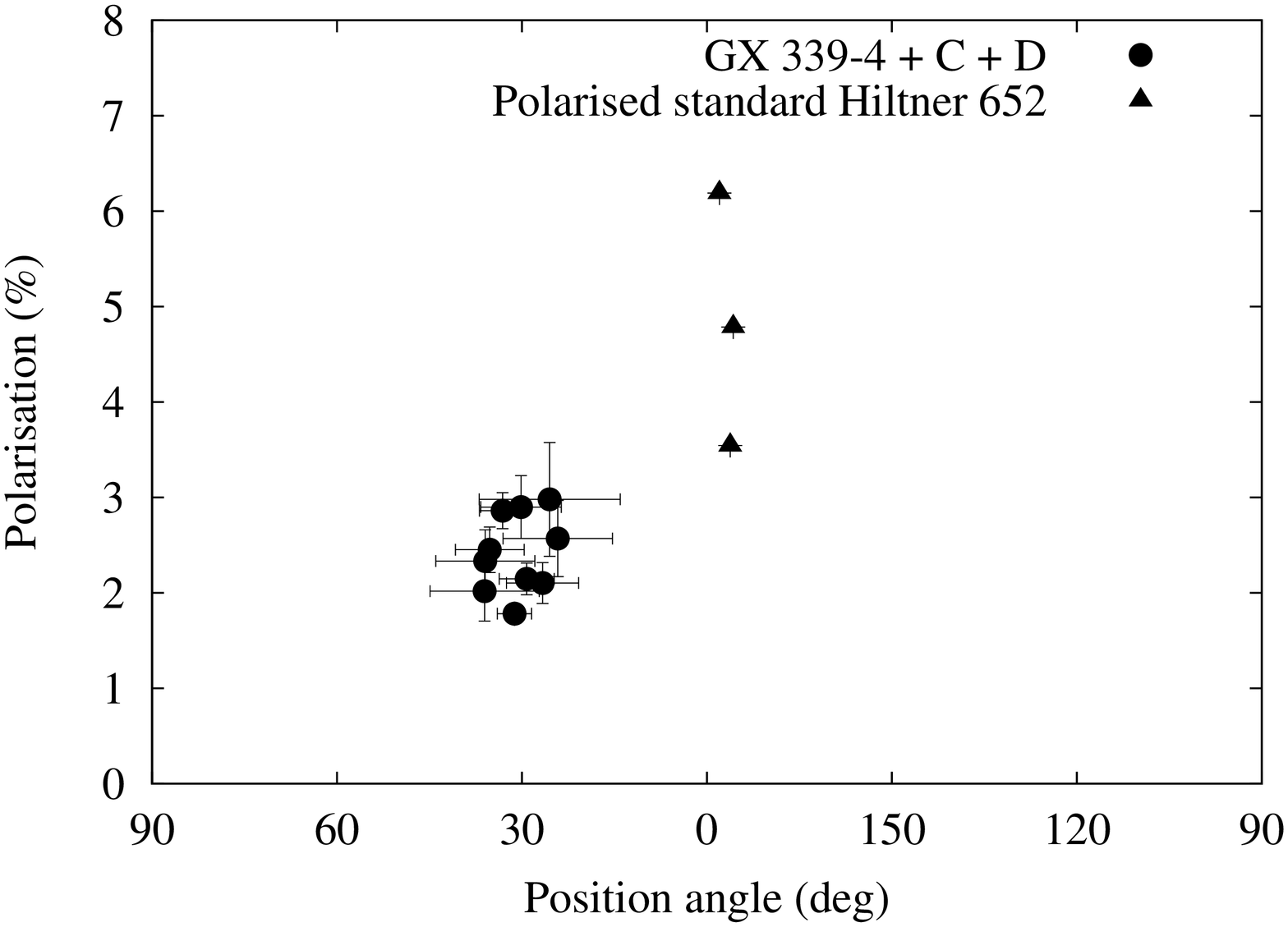}
\includegraphics[width=6cm,angle=270]{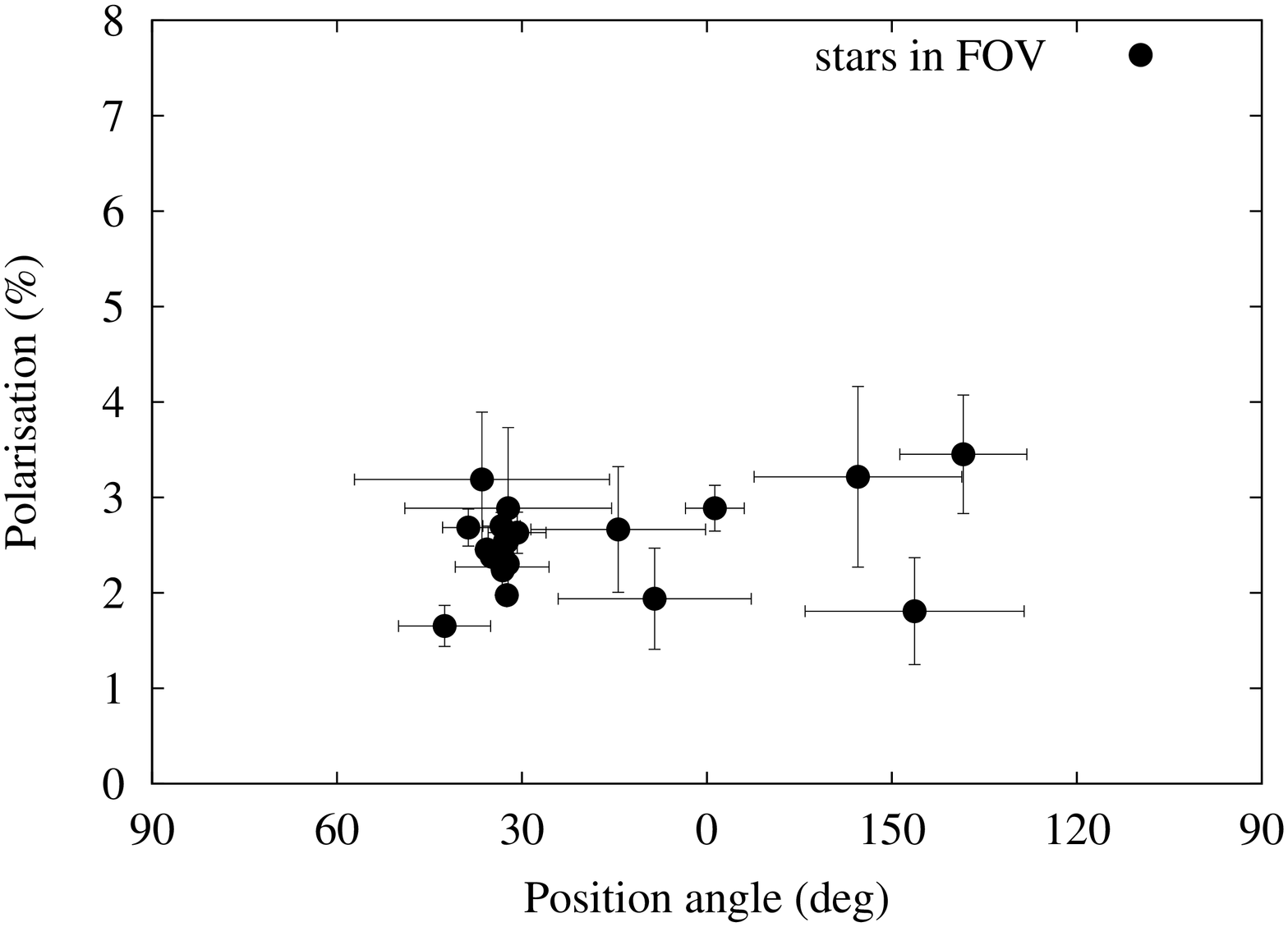}
\caption{Optical $V$, $R$ and $I$-band LP (when it is detected at the $> 3\sigma$ level) versus PA for GX 339--4 (including the two contaminating stars C and D) and and the polarised standard Hiltner 652 in the upper panel and three stars in the field of view in the lower panel. We plot a point for each observation in each filter. The polarisation of GX 339--4 is similar in level and PA to most of the data of the three field stars, and is likely to be caused by interstellar dust.}
\end{figure}

The magnitudes we obtain of GX 339--4 alone ($V\sim 18.7$, $R\sim 18.6$, $I\sim 17.5$; not including the contaminating stars) are similar to the $V$ and $I$ magnitudes measured a month or so earlier (C. Bailyn, private communication) but are $\sim$1.5 magnitudes brighter than the lowest level of $r=20.1$ \citep{shahet01}. Even at this lowest level the star contributes $\simlt 30$\% of the $r$-band emission \citep{shahet01}. The de-reddened optical SED of GX 339--4 for the five observations are shown in Fig. 3. The spectrum is blue and fairly steep between $R$ and $V$ ($\alpha = 2.1\pm 0.3$ between $I$ and $V$) indicating thermal emission, probably from the X-ray heated disc. However there is an apparent a flattening towards the $I$-band. This could be the same $I$-band flattening as seen during outburst, which is the optically thin jet spectrum beginning to dominate \citep{corbfe02,homaet05a}. The $R$-band flux errors are large however, and there may be no flattening. We may expect to see short timescale flux variability if the jet does play a role; from column X of Table 3 we see that GX 339--4 varies by $\simlt 5$\% in the high S/N observations, and the variability is not stronger in the $I$-band. We cannot therefore make any conclusions as to whether the jet component is polarised or not as the jet contribution to the $I$-band at these low flux levels is uncertain. We note that the optically thin radio jet ejecta seen from this source are polarised at a level of 4--9\% \citep{gallet04}. We do not plot a polarisation spectrum (LP versus $\nu$) for GX 339--4 because we find no relation between P and $\nu$ for GX 339--4 or for any of the field stars, within the errors of each measurement.
\newline\newline
\textbf{GRO J1655--40}:
\newline
$\sim 5$\% LP is detected in this source at the 3$\sigma$ level in $H$ and $Ks$, when it was in a hard state at the end of its 2005 outburst. In the upper panel of Fig. 4 we plot the polarisation spectrum, including the optical measurements of \cite{glioet98} during outburst in 1997. The optical polarisation, which varies as a function of orbital phase (taken during a soft X-ray state), is caused by local scattering and should decrease at lower frequencies. We see a statistical increase in LP in the NIR (shown by the fit to the optical--NIR data) which must therefore have a different origin to the optical LP. Our $Ks$-band LP of 5.8$\pm$1.9\% is not consistent with the 95\% confidence upper limit of LP$<$1.3\% in $Ks$ seen during quiescence \citep{dubuch06}. The NIR LP we measure must be intrinsic and transient, and we interpret it as originating from the optically thin region of the jets (see Section 3.2).

\begin{figure}
\centering
\includegraphics[width=6cm,angle=270]{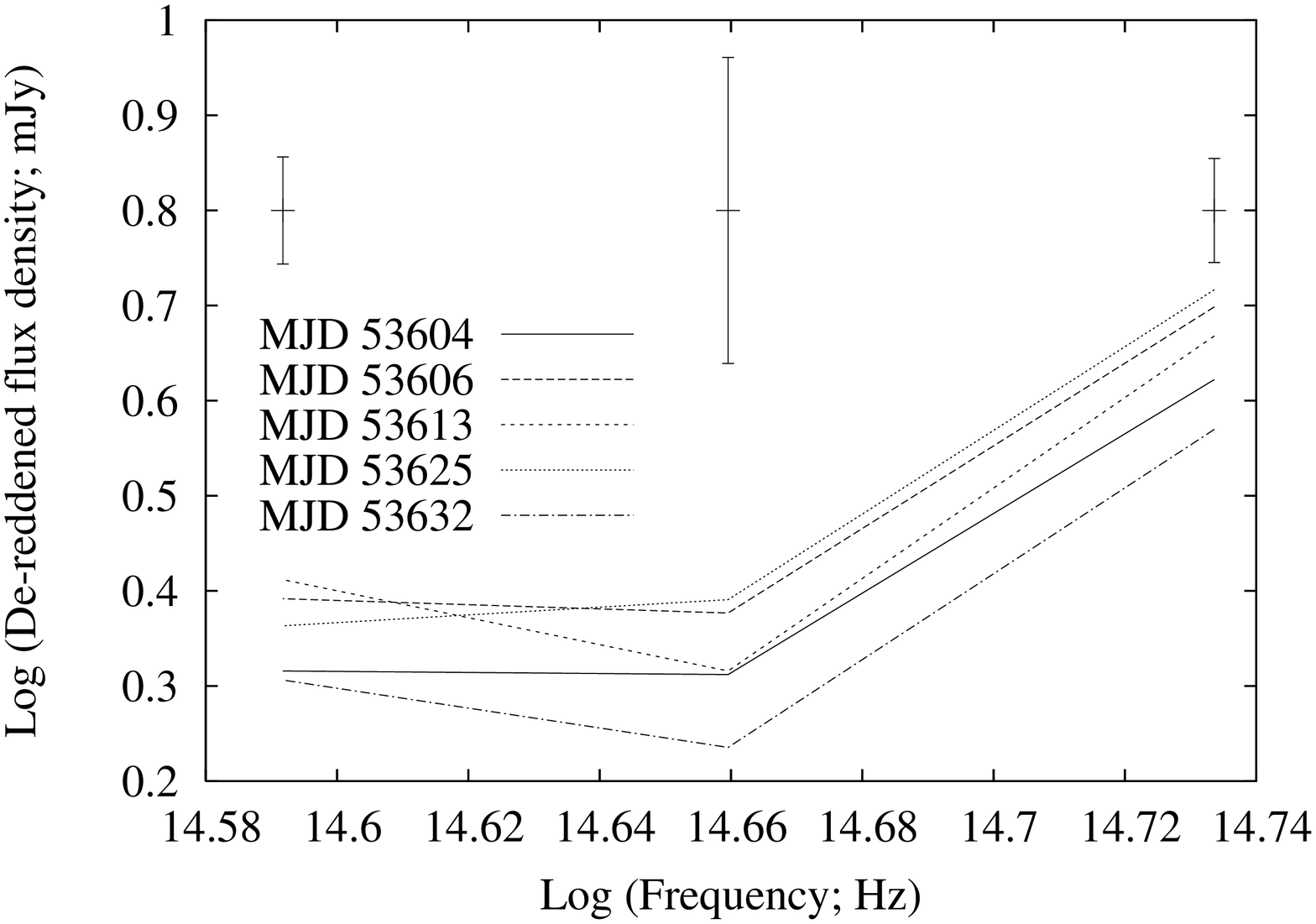}
\caption{The de-reddened $IRV$ spectrum of GX 339--4 (not including the two contaminating stars C and D). The error bars represent the mean error in flux density in each filter. The $I$-to-$V$ spectral index is $\alpha = 2.1\pm 0.3$.}
\end{figure}

The PA of the optical LP (120--130$^\circ$) is also different from that measured both in $H$ (44$\pm 18^\circ$) and $Ks$ (157$\pm18^\circ$). Rotation of PA with time is seen in some jets \citep[e.g.][]{hannet00,fendet02,gallet04} indicating an overall rotation of the local magnetic field. The $Ks$ LP detection is 14 days after the $H$, implying the magnetic field may be rotating during this time. The $H$-band PA is consistent with the direction of the jet on the plane of the sky; 47$^\circ$ \citep{hjelru95}, implying that the electric field vector is parallel to the jets at this time and the magnetic field is perpendicular to the jets. Two weeks later the field orientation had appeared to change by $\sim 70^\circ$. Interestingly, \cite{hannet00} found from well-sampled radio data that the field orientation changed only near the end of the 1994 outburst; here we see a change also at the end of its 2005 outburst.

The de-reddened $JHKs$ flux densities of GRO J1655--40 indicate a blue spectrum, and are similar to those measured by \cite{miglet07} between two and five weeks earlier. In their Fig. 6, the broadband SED shows the companion star dominating the optical and NIR, with the jet dominating only below $\nu \approx 5\times 10^{13}$ Hz. Our $JHKs$ observations are therefore also dominated by the star. According to the model in Fig. 6 of \citeauthor{miglet07} (upper panel), the jet contributes $\sim 40$\% and $\sim 30$\% of the emission in $Ks$ and $H$, respectively. If the jet is the source of the polarisation and the star and any other contributions are not polarised, this implies the jet itself is $15\pm 5$\% polarised in $Ks$ and $16\pm 5$\% polarised in $H$. We would not expect this level of LP from the optically thick jet, so the jet spectrum must turn over to become optically thin ($\alpha \approx -0.6$) at some frequency above $\sim 5\times 10^{13}$ Hz \citep[contrary to the models of][]{miglet07}. If we take the faintest possible NIR jet scenario, one where the turnover is at $5\times 10^{13}$ Hz and the optically thin spectrum is steep ($\alpha = -0.8$), the jet would contribute 14\% of the light at $Ks$ and 9\% at $H$. This would result in a jet LP of $42\pm 13$\% in $Ks$ and $52\pm 17$\% in $H$. The likely level of jet contribution is between the two above estimates, so a jet LP of 15--42\% in $Ks$ and 16--52\% in $H$ can explain the data. These results indicate a fairly ordered magnetic field, with $f = 0.41\pm 0.19$ from the $Ks$-band result (The $H$-band is less constraining).

\begin{figure}
\centering
\includegraphics[width=8.9cm,angle=0]{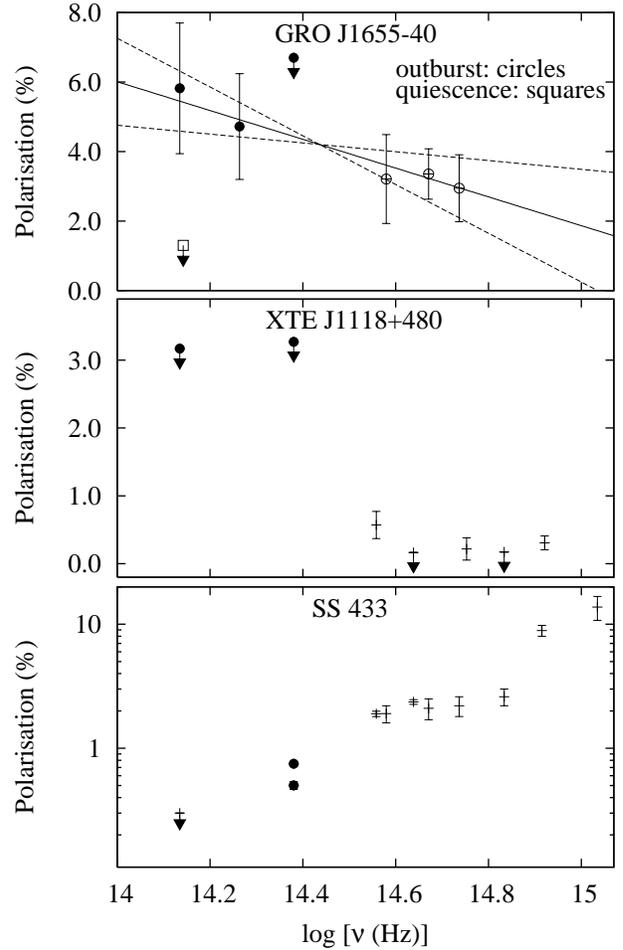}
\caption{LP as a function of frequency for three sources. Filled symbols are our results (upper limits are 1$\sigma$), other symbols are from the literature. $VRI$-band data of GRO J1655--40 are from \citet{glioet98} and a $K$-band upper limit in quiescence is from \citet{dubuch06}. The lines in the top panel show the fit to the outburst data and uncertainty in the slope. $UBVRI$ data of XTE J1118+480 are from \citet{schuet04}. $UV,UBVRI$ data of SS 433 are from \citeauthor{dolaet97} (1997; the contribution from interstellar LP has been subtracted from these data) with additional $RI$ data from \citet{schuet04} and a $K$ upper limit from \citet{thomet79}.}
\end{figure}

\begin{figure}
\centering
\includegraphics[width=8.9cm,angle=0]{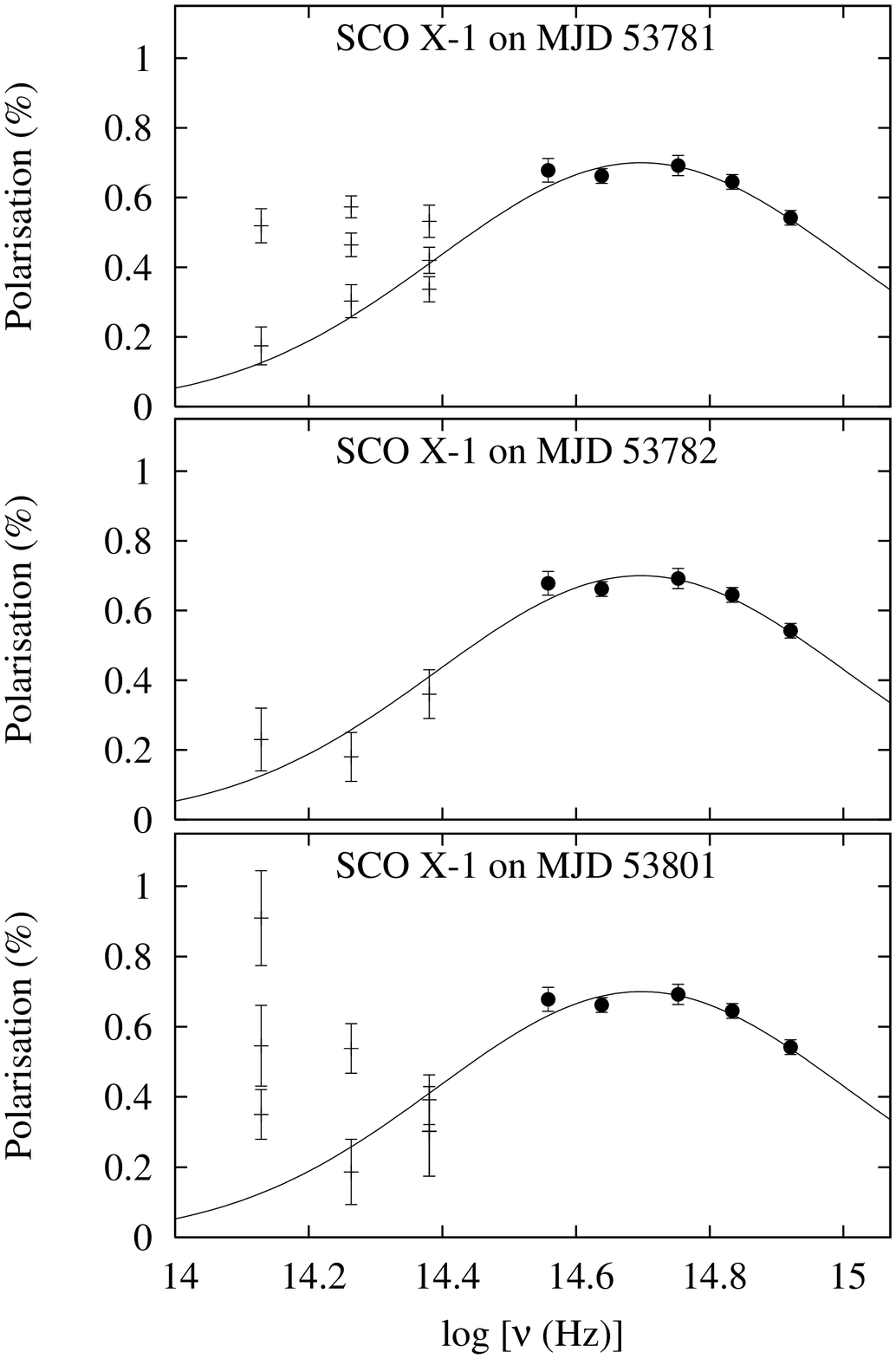}
\caption{LP as a function of frequency for Sco X--1 in three epochs. We include the optical data (filled circles) and the fit to this data for an interstellar origin to the polarisation (solid curve) from \citet{schuet04}. The crosses on the left are our $KHJ$-band results (2$\sigma$ detections of LP); a clear variable LP component above that expected from interstellar dust is seen at the lowest frequencies.}
\end{figure}

In addition, the flux from GRO J1655--40 is marginally variable in $Ks$ (7$\pm$3\% standard deviation; column X of Table 3) on short timescales ($\sim 20$ sec time resolution). If the star is not variable then this suggests the jet component is varying by 9--43\%. This is consistent with observations of other sources; large-amplitude variability has been seen associated with NIR optically thin synchrotron emission from the jets in XTE J1118+480 \citep{hyneet06}. We note however that strong variability can also be observed from other components, such as irradiation of the accretion disc.
\newline\newline
\textbf{GRO J0422+32}:
\newline
The apparent magnitudes and $J-K$ colour of this source are found to be consistent with the quiescent values measured by \cite{geliha03}. The S/N is low; we are able to place a 3$\sigma$ upper limit on the $J$-band polarisation; LP$<$11.6\%. It has been shown that the star likely dominates the NIR light \citep{geliha03} although recent observations detect a strong flickering component which may come from the disc \citep*{reynet07}.
\newline\newline
\textbf{4U 0614+09}:
\newline
We detect this NSXB with a $K$-band magnitude $\sim 0.5\pm 0.2$ brighter than reported from observations made in 2002. The system is an ultra-compact NSXB whereby the white dwarf donor has not directly been detected. \cite{miglet06} showed that the mid-IR spectrum is dominated by optically thin synchrotron emission from the jets (with $\alpha = -0.6$), and this joins the disc component ($\alpha \approx +2$) around the $H$-band. Therefore our $K$-band measurement is likely to be jet-dominated. Despite this we do not detect LP, with a 3$\sigma$ upper limit of 16\%. According to the spectrum in \cite{miglet06}, the jet contributes around twice as much light as the disc in the $K$-band. Assuming that the accretion disc contributes less than 40\% of the $K$-band light, the jet component must be less than 27\% polarised. The corresponding magnetic field ordering is $f < 0.38$.
\newline\newline
\textbf{XTE J1118+480}:
\newline
The polarisation spectrum of this source, including optical data from \cite*{schuet04} is shown in the centre panel of Fig. 4. \citeauthor{schuet04} claim the LP is variable (with a positive detection in some optical bands but not in others) and their strongest detection (2.6$\sigma$) is in the $I$-band; LP = 0.57$\pm$0.22\%. There is no apparent increase at higher frequencies, with a $B$-band upper limit of 0.38\% (3$\sigma$) so the positive detection is unlikely to be due to interstellar dust (the extinction is very low; see Table 2). We find 3$\sigma$ upper limits on the order of 7\% LP in $J$ and $K$. According to a recent study of the broadband quiescent spectrum of XTE J1118+480 \citep[using for the first time optical, NIR and mid-IR data;][]{gallet07}, the jet could make a significant contribution to the mid-IR--optical spectrum. In their model (middle right panel of their Fig. 3), the jet contributes approximately 35\%, 25\% and 28\% of the flux in $K$, $J$ and $I$, respectively. From the polarisation measurements, this would correspond to a jet LP of $< 21$\% ($K$), $< 25$\% ($J$) and 2.0$\pm$0.8\% ($I$). This is by no means a solid result, but the $I$-band detection suggests the jet does indeed contribute. The accretion disc is also likely to make a contribution to at least the optical wavebands \citep{mikoet05} but other origins of LP such as scattering are likely to be stronger at higher frequencies, contrary to observations \citep{schuet04}.
\newline\newline
\textbf{Sco X--1}:
\newline
The optical light of Sco X--1 is polarised at a low level due to interstellar dust \citep{landan72,schuet04} but there is one claim that it is variable and therefore intrinsic \citep{shakef75}. Recent NIR spectropolarimetry has revealed intrinsic LP in the $H$ and $K$ region of the spectrum, with a clear increase in LP below 1.4$\times 10^{14}$ Hz that is interpreted as coming from the (optically thin) jet \citep{shahet07}. Here, we report $JHK$ polarimetry over four epochs that shows a clear variability of LP which is stronger at lower frequencies (Fig. 5). In this figure, we include the optical data and interstellar model fit from \cite{schuet04} \citep[much like Fig. 3 of][]{shahet07}. The NIR LP cannot be explained by the interstellar model because it is significantly greater than that expected, and variable. The $J$-band data are all consistent with the model, whereas only two of the six $K$-band measurements agree.

\begin{figure}
\centering
\includegraphics[width=8.9cm,angle=0]{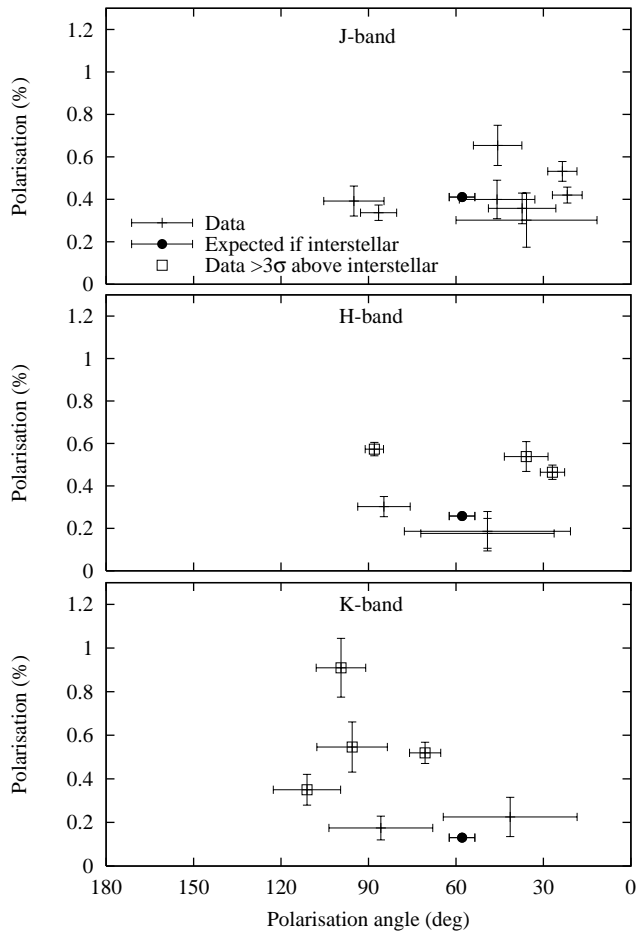}
\caption{The level of LP in Sco X--1 versus the position angle. Square symbols indicate data with LP level $>3\sigma$ above that expected from interstellar dust \citep{schuet04}. The axis of the resolved jet of Sco X--1 is at PA = $54^\circ$ \citep{fomaet01}.}
\end{figure}

Z-sources like Sco X--1 have the signature of optical/NIR colours and correlated (with X-ray) variability typical of an irradiated accretion disc \citep[e.g.][]{mcgoet03,russet07}. They also have a radio counterpart, likely from the jet \citep[e.g.][]{miglfe06}\footnote{Some interpretations prefer an origin located in the neutron star magnetosphere \citep[e.g.][]{paizet06}, but in Sco X--1 the jet is clearly resolved \citep*{fomaet01}.}, the spectrum of which may extend to the NIR, but is generally fainter than the disc in this regime \citep{russet07}. \cite{wachet06} have identified the jet component in Sco X--1 in the mid-infrared; the bright disc dominates at higher frequencies. The variable $K$ and $H$-band LP strongly suggest the jet is present (see Section 3.2 for discussions of alternative origins). The low level (maximum LP $\sim$ 1.0\%) does not rule out optically thick emission, which can produce this level. Our LP levels largely agree with those of \citeauthor{shahet07}; at the $K$-band central frequency they measure LP = 0.5--1.3\% whereas our levels range from LP = 0.13--0.91\%, and at $H$ they measure LP = 0.2--0.7\% and we have LP levels between 0.07\% and 0.57\%. \citeauthor{shahet07} show the LP could increase to up to $\sim 30$\% in the mid-IR. Our results reveal the LP to be highly variable and sometimes absent, suggesting either the conditions in the inner regions of the jet are variable, or the jet luminosity (and hense LP) is changing with time, perhaps correlated with the Z-track.

Interestingly, the PA of the most significant LP detections is $\sim 90$-$100^\circ$, differing from a PA of $\sim 60^\circ$ when the level is consistent with the interstellar model (Fig. 6). $60^\circ$ is also the PA of the interstellar polarisation measured in the optical, further confirming the origin of the NIR LP is likely to be interstellar when its level is consistent with the model. The resolved radio jets \citep{fomaet01} are in a direction (which is constant over many years) on the plane of the sky $\sim 40^\circ$ different from the PA of the polarised NIR jet. There may be systematic errors in our measurements of PA (Section 2.3) but the fact that our low-LP measurements of this source are consistent with the interstellar model suggest these errors are low. \cite{shahet07} measured a slightly different PA to us; 116--147$^\circ$ which is perpendicular to the jet axis, indicating a magnetic field aligned with the jet.

The total flux of Sco X--1 (which is uncalibrated) is generally variable by $< 1$\%. Any jet component is likely to be variable so this suggests a low jet contribution. If this is the case, the jet must be $>> 1$\% polarised, and so it must be coming from the optically thin region. It would be interesting to use simultaneous X-ray data in a future study, to test whether the LP (and flux) is correlated with the position on the Z-track.
\newline\newline
\textbf{Aql X--1}:
\newline
An optical polarimetric study of this source was carried out by \cite{charet80} who found LP $\sim$ 1.5\%, which they interpret to be interstellar in origin. However they also detected maginal variability, with one measurement of LP = 2.3$\pm$0.5\%. The S/N was low in our NIR observations and we can place 3$\sigma$ upper limits of LP $<$ 5\%, 17\% and 12\% in $J$, $H$ and $K$, respectively \citep[the aperture includes a very close contaminating star which appears to contribute about half the flux in $K$; star $a$ in][]{chevet99}.

Our observations span three epochs, the first of which was at the end of a small outburst \citep*{bailet06}. Indeed, we derive $JHK$ magnitudes which are consistent with those found in quiescence by \cite{garcet99} except on our first epoch (20 Aug 2006), where $J$ and $K$ are 0.20$\pm$0.15 and 0.27$\pm$0.16 magnitudes brighter than the mean quiescent values. On this date, the $J$--$K$ de-reddened SED is blue (column VI of Table 4) implying the emission is from the disc, although the contaminating star will affect this measurement. The disc and donor star are expected to dominate the optical/NIR light in quiescence, but during outburst the contributions vary. A dramatic reddening of the optical SED was seen at the peak of the bright 1978 outburst of Aql X--1 \citep{charet80} which could be explained by the domination of the jet component \citep{russet07}. However the jet is likely to contribute very little to the optical/NIR light in quiescence.
\newline\newline
\textbf{SS 433}:
\newline
High levels of LP have been detected in this high-mass X-ray binary in the optical and UV, which can be explained by a combination of Thompson and Rayleigh scattering \citep{dolaet97}. In the lower panel of Fig. 4 we show the polarisation spectrum of SS 433 with our two $J$-band measurements, the optical/UV data and a $K$-band upper limit from \cite{thomet79}. Our data follow the general trend of decreasing LP with decreasing frequency, so we expect the origin is the same as in the optical. However the PAs differ between the $J$-band ($70-80^\circ$ during two epochs) and the optical ($\sim 90-140^\circ$), which could be explained by an interstellar contribution at these low levels of LP (the extinction is high; see Table 2).

\subsection{On the origin of the variable NIR polarisation in GRO J1655--40 and Sco X--1}

Many phenomena could result in polarised NIR emission from LMXBs. Here we discuss the likely causes of LP and use the GRO J1655--40 and Sco X--1 results to constrain the origin of the polarisation.
\newline\newline
\textbf{Thompson or Rayleigh scattering}:
\newline
Scattering of light within the system can result in detectable LP (see Section 1); the optical light from A0620--00 and GRO J1655--40 is polarised in this way at a level of a few percent \citep{dolata89,glioet98}. However, LP due to scattering is expected to decrease as a function of increasing wavelength and not be more than a few percent in the infrared \citep*{browet78,dola84}. The 5--6\% we measure for GRO J1655--40 therefore cannot be explained by local scattering, and in both this source and Sco X--1 the transient NIR LP is greater than the optical LP. In high-mass X-ray binaries (HMXBs) the scattering of stellar light can produce LP \citep[e.g.][]{kempet83}. In Sco X--1 this cannot be the case as the accretion disc dominates the optical and NIR and as such any polarised stellar light would be diluted by the light from the disc \citep{mcnaet03,wachet06,russet07}. For GRO J1655--40 the star is large and bright and peaks in the optical \citep{miglet07} but the high level of NIR LP and the negative relation between LP and frequency allow us to rule out scattered stellar light as the origin of the NIR LP.

Scattering also produces LP that is variable on timescales of the orbital period; this has been significantly detected from optical observations of both HMXBs and LMXBs \citep[e.g.][]{kempet83,dolata89,glioet98}. However, Fig. 5 shows that the amplitude of the NIR polarisation of Sco X--1 varies dramatically between three consecutive observations on the same date (within 30 minutes), which is very different from the 18.9 hour orbital period of Sco X--1. The combination of these lines of evidence strongly indicates we can rule out scattering as the origin to the transient LP in both GRO J1655--40 and Sco X--1.
\newline\newline
\textbf{The scattering of jets on the companion star surface}:
\newline
If the jet does not escape the system at an angle approximately perpendicular to the plane of the disc, LP may be caused by direct interaction between the jet plasma and the companion star surface. In LMXBs however, the companion typically occupies $\sim 3$\% of the solid angle as seen from the compact object \citep[e.g. see Fig. 1 of][]{obriet02}. In this case, a jet colliding with the star requires either a jet opening angle of $\sim 70^\circ$ \citep[which contradicts the morphology of the resolved radio jets in both sources;][]{tinget95,fomaet01} or would have to be directed towards the companion at an angle away from the disc of $< 20^\circ$, as opposed to the more likely $\sim 90^\circ$. The two NIR LP detections of GRO J1655--40 are different in orbital phase by 0.3 \citep[they are 14.0 days apart and the orbital period is 2.62 days;][]{jonket04} and in Sco X--1 the transient LP is measured at many orbital phases (Table 4), so this latter scenario can be ruled out.
\newline\newline
\textbf{Synchrotron emission from the inner regions of the jets}:
\newline
Optically thin synchrotron emission is the only process in LMXBs capable of producing NIR LP of levels greater than a few percent and can, depending on the structure of the magnetic field, produce tens of percent of LP. In  both sources, the jet is known to become dominant at wavelengths just longer than the NIR \citep{wachet06,shahet07,miglet07} so light from the jet is known to contribute significantly at NIR wavelengths (we constrain the level of jet contribution for GRO J1655--40 in Section 3.1). Optically thin synchrotron emission is the only mechanism in our systems capable of producing the high level of NIR LP in GRO J1655--40 since we have ruled out local scattering. In addition, synchrotron emission (both optically thick and thin) is likely the only process that can account for the fast flickering of LP in Sco X--1. \cite{shahet07} also rule out alternative origins to the NIR LP in Sco X--1. Moreover, we would expect LP from the jet component to be weaker at shorter wavelengths, as is observed in both sources, since the disc (of Sco X--1) and companion star (of GRO J1655--40) are much more luminous than the jet at shorter wavelengths.

From the lines of evidence in this Section, we can conclude with confidence that the origin of the LP in these two sources is very likely synchrotron emission from the inner regions of the jets.

\section{Summary and conclusions}

We have presented linear polarimetric observations of seven X-ray binaries in the NIR regime and one in the optical. In one black hole system and one neutron star system (GRO J1655--40 and Sco X--1), intrinsic LP is detected at the $> 3\sigma$ level, which is stronger at lower frequencies. We discuss possible origins of the LP and conclude that it is very likely the signature of NIR synchrotron emission from their jets, which must reside in partially-ordered magnetic fields. We rule out scattering within the system as the origin of the LP. The intrinsic LP in Sco X--1 is at a comparable level to that measured recently by \cite{shahet07}, but we find it is also highly variable in $H$ and $K$-bands; sometimes it is absent. The variability could be due to changes in either the jet magnetic field structure or the jet power and luminosity. SS 433 also possesses intrinsic NIR LP which is likely due to local scattering. The optical light from GX 339--4 is polarised, almost certainly due to interstellar dust.

We have been able to constrain $f$, the dimensionless parameter representing the ordering of the magnetic field in the region of the jets, in two sources. In the BHXB GRO J1655--40, $f = 0.41\pm 0.19$ whereas in the NSXB 4U 0614+09, $f < 0.38$. The BHXB result is consistent with those found from radio studies of the transient jet ejections in BHXBs \citep[e.g.][]{fendet02,brocet07} and multi-wavelength studies of jets in AGN \citep[e.g.][]{saiksa88,jorset07}. The inner regions of steady jets from BHXBs may have a similar magnetic field configuration to the brighter, transient jets. A dramatic change in PA in GRO J1655--40 at the end of its outbursts (also seen in the radio) suggests the magnetic field structure may change at low jet powers, and/or the polarisation may originate from a different region of the jet at these later times. For NSXBs, this is the first time a constraint on $f$ has been made (none exist from radio observations). The non-detection of polarisation in any of our low-luminosity sources suggests either the jet contribution is insignificant or the magnetic field is not well-ordered.

We have demonstrated that polarimetry is a powerful tool to constrain the conditions in the inner regions of the jets in X-ray binaries. The NIR region of the spectrum of LMXBs is usually dominated by optically thin synchrotron emission from the jets, at least in outburst in the hard state (e.g. Russell et al. 2006; 2007). Future polarimetric studies may reveal changes in the conditions as a function of jet power (i.e. jet luminosity) and differences between object types (black hole or neutron star systems). In particular, the BHXB GX 339--4 undergoes quasi-regular outbursts with a bright NIR jet \citep[e.g.][]{homaet05a}. In quiescence, a positive polarisation detection from the jet will indicate its contribution is significant (supporting jet-dominated scenarios), and therefore some estimates of the mass function, which often do not consider jet contributions, may be inaccurate.

\vspace{5mm}
\emph{Acknowledgements}.
We thank Tariq Shahbaz for copies of his paper \citep{shahet07} prior to publication. In particular it was useful to compare their Sco X--1 results to ours. DMR would like to thank Stephane Corbel and Christian Knigge for stimulating discussions. We acknowledge Charles Bailyn for providing optical magnitudes of GX 339--4 near in time to our observations. Based on Service Programme observations made with the United Kingdom Infrared Telescope (UKIRT), which is operated by the Joint Astronomy Centre on behalf of the Science and Technology Facilities Council of the U.K. We thank the Department of Physical Sciences, University of Hertfordshire, for providing IRPOL2 for the UKIRT. Also based on observations collected at the European Southern Observatory, Chile, under ESO Programme IDs 076.D-0497 and 275.D-5062.

\end{document}